\documentclass[twocolumn]{aastex631}

\begin{document}

\title{HybPSF: Hybrid PSF reconstruction for the observed JWST NIRCam image}

\correspondingauthor{Huanyuan, Shan}
\email{hyshan@shao.ac.cn}

\author[0000-0001-6495-1890]{Lin, Nie}
\affiliation{Department of Information Engineering,
Wuhan Institute of City,
Wuhan, Hubei 430083, China}

\author[0000-0001-8534-837X]{Huanyuan,Shan}
\affiliation{Shanghai Astronomical Observatory (SHAO), 
Nandan Road 80, 
Shanghai 200030, China }
\affiliation{Key Laboratory of Radio Astronomy and Technology, Chinese Academy of Sciences, \\
A20 Datun Road, Chaoyang District, Beijing, 100101, P. R. China }
\affiliation{University of Chinese Academy of Sciences, Beijing 100049, China}

\author{Guoliang, Li}
\affiliation{Purple Mountain Observatory, 
Chinese Academy of Sciences, 
Nanjing, Jiangsu, 210023, China}

\author{Lei, Wang}
\affiliation{Purple Mountain Observatory, 
Chinese Academy of Sciences, 
Nanjing, Jiangsu, 210023, China}

\author[0000-0003-0202-0534]{Cheng, Cheng}
\affiliation{Chinese Academy of Sciences South America Center for Astronomy,\\ National Astronomical Observatories, CAS, Beijing 100101, China}
\affiliation{CAS Key Laboratory of Optical Astronomy, National Astronomical Observatories, \\
Chinese Academy of Sciences, Beijing 100101, China}

\author{Charling, Tao}
\affiliation{Department of Astrophysics,
Tsinghua University, 
Beijing 100084, 
China}
\affiliation{
Aix Marseille Université,
CNRS/IN2P3,
CPPM UMR 7346,
F-13288 Marseille,
France
}

\author{Qifan, Cui}
\affiliation{Key Laboratory of Space and  Technology, National Astronomical Observatories, \\
Chinese Academy of Sciences, Beijing 100101, China}

\author{Yushan, Xie}
\affiliation{Shanghai Astronomical Observatory (SHAO), 
Nandan Road 80, 
Shanghai 200030, 
China}

\author{Dezi, Liu}
\affiliation{South-Western Institute for Astronomy Research, Yunnan University, Kunming, Yunnan, 650500, China}

\author{Zekang, Zhang}
\affiliation{Shanghai Astronomical Observatory (SHAO), 
Nandan Road 80, 
Shanghai 200030, 
China}
\affiliation{
University of Chinese Academy of Sciences, 
Beijing 100049, 
PR China
}



\begin{abstract}

The James Webb Space Telescope (JWST) ushers in a new era of astronomical observation and discovery, offering unprecedented precision in a variety of measurements such as photometry, astrometry, morphology, and shear measurement. Accurate point spread function (PSF) models are crucial for many of these measurements. In this paper, we introduce a hybrid PSF construction method called HybPSF for JWST NIRCam imaging data. HybPSF combines the WebbPSF software, which simulates the PSF for JWST, with observed data to produce more accurate and reliable PSF models. We apply this method to the SMACS J0723 imaging data and construct supplementary structures from residuals obtained by subtracting the WebbPSF PSF model from the data. Our results show that HybPSF significantly reduces discrepancies between the PSF model and the data compared to WebbPSF. Specifically, the PSF shape parameter ellipticity and size comparisons indicate that HybPSF improves precision by a factor of approximately 10 for $R^2$ and $50\%$ for $e$. This improvement has important implications for astronomical measurements using JWST NIRCam imaging data.

\end{abstract}



\section{Introduction} \label{sec:intro}
The JWST provides us an unprecedented chance to gaze our universe. 
With its powerful observation capability in near infrared wavelengths, more details about distant objects could be observed and variety properties of the universe at high-redshift can be investigated more precisely. 

We are interested in galaxy clusters which are the largest gravitationally bounded structures in the universe. By studying the mass distribution of galaxy clusters, we can learn more about the dark matter clustering properties \citep{2008ApJ...681..187B,2011A&ARv..19...47K,2012ApJ...748...56S,2018NatAs...2..744S,2022A&A...664A..90L}, the investigations of star formation efficiency in cluster contexts by comparing the distribution of the ratio of stellar mass to total mass within galaxy clusters\citep{1995ApJ...447L..81B,2012MNRAS.426.3369J,2014MNRAS.439.2505B,2016ApJ...831..182H,2016MNRAS.457.2029J,2018ApJ...859...58F}. Studying the high redshift universe with the help of magnification effect of clusters\citep{2018ApJ...868..129S,2022Natur.603..815W,2023ApJ...946L..35M,2023ApJS..264...15S}. And the nature of dark matter(such as the  self-interaction cross-section \citep{2004ApJ...606..819M,2008ApJ...679.1173R,2018ApJ...869..104W,2019MNRAS.488.1572H,2021MNRAS.507.2432G,2022MNRAS.510...54A}) could also be investigated from the spatial offset of the baryonic and dark-matter profiles in clusters. 

Accurate mass distribution measurements of galaxy clusters require combining both strong and weak lensing signals in mass modeling methods \citep{2004astro.ph.10643B,2005A&A...437...39B,2006ApJ...652..937B,8173886,2016ApJ...821..116U,2017ApJ...844..127W,Congdon2018}. In the weak lensing regime, the precision of shear measurements determines the accuracy of mass distributions, and one of the most significant systematic errors comes from inaccurate Point-Spread-Function (PSF) modeling \citep{2005astro.ph..9252S,schneider2006weak,2018ARA&A..56..393M}. The PSF describes how point sources would appear in the image and smears the intrinsic galaxy image, ultimately affecting the accuracy of galaxy shape measurements. The PSF is mainly affected by atmospheric turbulence (for ground-based telescopes), telescope optics, and detector effects.
The accuracy of shear measurements is heavily affected by imprecise PSF modeling, and reliable mass modeling for galaxy clusters also requires credible PSF modeling. Additionally, astrometry and photometry of point sources cannot be done with high precision without PSF-fitting \citep{2000PASP..112.1360A,2005MNRAS.361..861M,2022MNRAS.517..484N}. Therefore, accurate PSF modeling is crucial for various astronomical measurements and analyses, including mass distribution measurements of galaxy clusters, shear measurements, astrometry, and photometry.

Generally, PSFs should be first measured at star positions and then properly interpolated to galaxy positions \citep{2018ARA&A..56..393M,2021MNRAS.501.1282J,2018MNRAS.481.1149Z}, and a certain number of stars are usually required for PSF modeling. However, for space telescopes like HST and JWST, star images are often undersampled \citep{1989ESOC...31..215A,1999PASP..111..227L,2022ApJ...940L...7W,2009SPIE.7436E..0DS}, and the number of PSF stars in a single exposure might not be enough to construct the PSF model for classical PSF modeling methods \citep{2002AJ....124.3255V,2007ApJS..172..203R,2020MNRAS.496.5017G}. Therefore, forward modeling methods (e.g., Tiny Tim for HST \citep{2011SPIE.8127E..0JK} and WebbPSF\footnote[1]{\url{https://webbpsf.readthedocs.io/en/latest/intro.html}} \citep{2012SPIE.8442E..3DP,2014SPIE.9143E..3XP,2015ascl.soft04007P} for JWST) are essential for obtaining the PSF of observed images.Forward modeling methods can model the PSF by simulating PSFs for a telescope's instruments in imaging modes. However, there may be discrepancies between the model and data due to differences in instrument configurations between simulations and practical telescope status \citep{2018AAS...23115036H}. To obtain more accurate PSF models, some improved methods have been developed \citep{2007ApJS..172..203R,2012ApJS..203...24V,2019AAS...23324519B,2021A&A...646A..27L,2023InvPr..39c5008L}. For example, WaveDiff \citep{2023InvPr..39c5008L} reconstructs the best fit wavefront based on a differentiable optical forward model according to the data to obtain better PSF models. WebbPSF calculates the PSF by using the results of in-flight wavefront sensing measurements (Optical Path Difference files, OPD), which are produced by the JWST Wavefront Sensing Subsystem (WSS) roughly every two days in flight \citep{2019AAS...23324519B}. However, there is still a prominent discrepancy at the inner region of the PSF between data and WebbPSF models \citep{2022ApJ...939L..28D,2022arXiv220813582O}.

In this paper, we introduce a hybrid PSF modeling method, HybPSF, for JWST NIRCam images that combines simulated PSFs from WebbPSF with observed data. We analyze the residuals obtained by subtracting the WebbPSF PSF model from the data using Principal-Component-Analysis (PCA) method and construct the residual model in the images. PCA is a statistical non-parametric procedure that converts a set of observations of possible variables into a set of linearly uncorrelated principal components and is popular for data compression and information extraction \citep{pearson1901liii,2014arXiv1404.1100S,2016RAA....16..139L,2021MNRAS.503.4436N}. The final PSF model is a composition of the WebbPSF model and the constructed residual model. SMACS J0723 is a massive galaxy cluster (z$\sim$0.388) \citep{2001ApJ...553..668E,2018MNRAS.479..844R} previously observed by the Hubble Space Telescope (HST) in the HST/ACS and WFC3 cameras \citep{2019ApJ...884...85C}. JWST's First Deep Field has released observed data of the massive galaxy cluster SMACS J0723 \citep{2022ApJ...936L..14P}, which were performed with NIRCAM filters F090W, F150W, F200W, F277W, F356W, and F444W. We will use the SMACS J0723 NIRCam imaging data to demonstrate our method. The advantage of SMACS J0723 is that there are more stars, which are from LMC \citep{2023arXiv230613037S}, that can be used in PSF modeling and facilitate accurate shear measurement.

This paper is structured as follows: We introduce the data reduction for star image selection and the discrepancy between the model and data in Section \ref{sec:image} and \ref{sec:discrepancy} respectively. The hybrid PSF model and its comparisons with WebbPSF model is described in Section \ref{sec:hybrid}. We provide a summary of this work in Section \ref{sec:summ}.

\section{data reduction and star image selection}
\label{sec:image}

In this work, we use the JWST first released NIRCam Imaging data of SMACS0723, including F090W, F150W, F200W in the short wavelength channel and F277W, F356W, and F444W in the long wavelength channel, to develop and validate our PSF modeling method. We started our data processing based on the "stage two" products, which are processed by JWST\footnote[2]{\url{https://jwst-pipeline.readthedocs.io/en/stable/jwst/introduction.html}} version 1.8.1 in the context of $jwst\_1017.pmap$ \citep{2017ASPC..512..355B,2019ASPC..523..543B,2022zndo...7038885B}. The "1/f" noise \citep{1965PASP...77..133L,2015jwst.rept.4118R} is also removed using a method developed by Wang et al. (in prep.)\footnote[3]{\url{https://github.com/chriswillott/jwst}}. We mark cosmic rays or bad pixels using an improved statistical algorithm in multiple exposures coaddition to reduce their impact on PSF reconstruction. To minimize the effect of cosmic rays or bad pixels removal on PSF reconstruction, we fill normal flux values to the marked pixels generated by the statistics over the normal coadded pixels on the same position. Finally, we extract our PSF star images from the above processed data and convert pixel values into counts \citep{2022MNRAS.517..484N}.

\begin{figure}[ht!]
\centering
\includegraphics[width=.5\textwidth,clip]{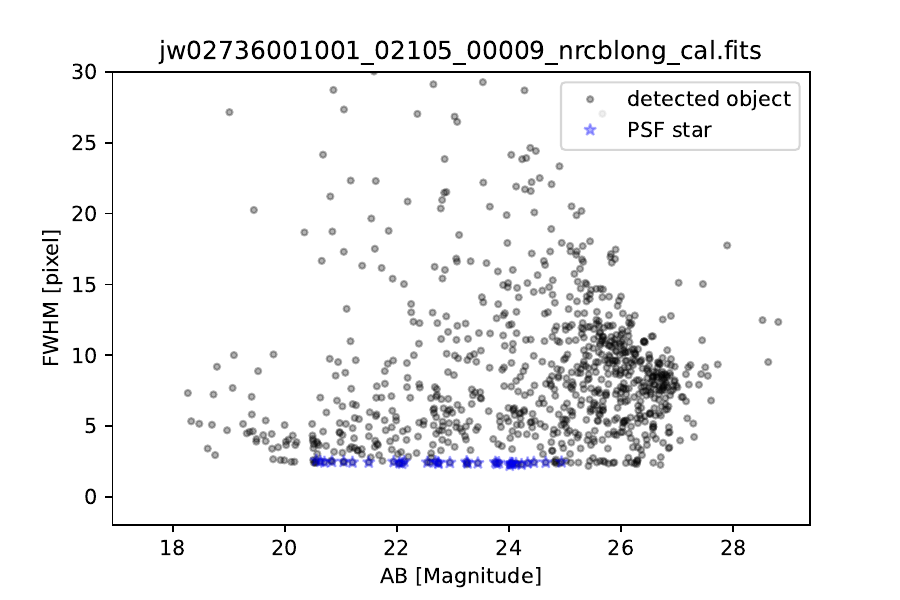}
\caption{The distribution ``FWHM'' as a function of ``magnitude'' for all object detected from ``$jw02736001001\_02105\_00009\_nrcblong\_cal.fits$'' by SExtractor. The ``AB [Magnitude]'' is estimated from: $FLUX\_AUTO$ measuring from SExtractor, and ``FWHM [pixel]'' corresponds to the value of ``$FWHM\_IMAGE$'' from SExtractor. The blue stars at the bottom of image represent the PSF stars.}
\label{star_locas}
\end{figure}

We select our PSF star image candidates according to the ``FLUX$\_$RADIUS", ``FWHM$\_$IMAGE" and ``FLUX$\_$AUTO" parameters measured from SExtractor \citep{1996A&AS..117..393B,2010ascl.soft10064B}. In the parameter space, stars appear as a locus of points with constant size and are separate from the larger cluster of galaxies \citep{2021MNRAS.501.1282J}. Figure \ref{star_locas} shows an example of the selected PSF stars from channel F444W, ``$jw02736001001\_02105\_00009\_nrcblong\_cal.fits$'' in the parameter space where the star locus appears separate from the other points. We use the Friends-of-Friends (FOF) method to analyze the images and select stamps that have multiple sources as blended stamps \citep{2019ApJ...875...48Z}.
%Each source is identified as a number of connected pixels by using the FOF. The number of pixels of each source should be larger than 7 and pixel value is required larger than 2$\sigma$ of the background. 
Each source is identified as a number of connected pixels that are above 2$\sigma$ by using the FOF method, and the number of pixels in each group should be larger than 7. We found that over $20\%$ and $80\%$ of the stamps in short and long wavelength channels are blended, respectively.
Next, we select unblended stamps as PSF star image candidates. For the blended stamps, we choose those stamps with the disconnected pixels of multi objects as  candidates. To avoid the effect of bright pixels from non-star objects, the pixels of objects located at the edge of the stamps are masked and set to zero. Meanwhile we construct a mask image $\mathcal{M}$ for each stamp in which the masked pixels have zero value and the others are set to 1 for each selected blended image. 

\begin{figure}[ht!]
\centering
\includegraphics[width=.48\textwidth,clip]{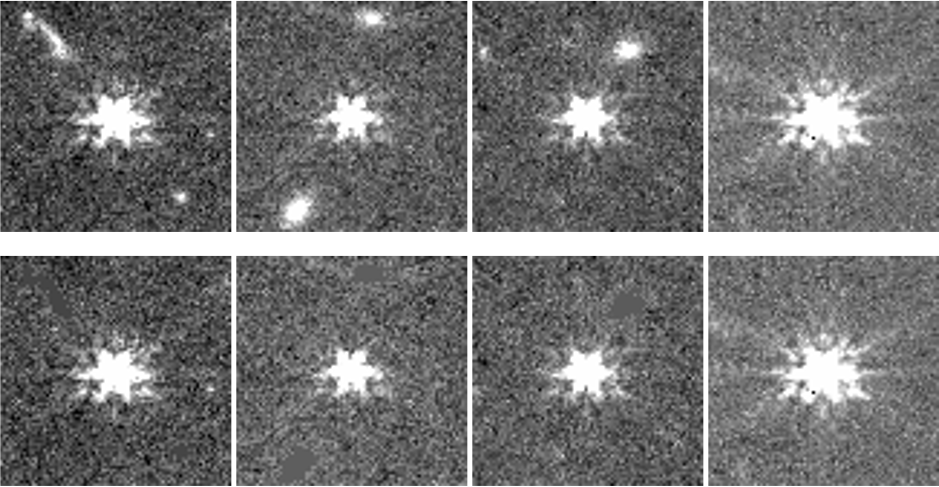}
\caption{Examples of our selected star stamps. The upper panels represent the star image candidates without masking the neighboring sources, and the lower panels shows the corresponding images after masking.}
\label{star_stamps}
\end{figure}

Besides, faint star images generally are inappropriate for PSF modelings \citep{2009A&A...500..647P,2022MNRAS.tmp.3141Z}. In order to use as many stars as possible for PSF modeling, we choose stamps that have SNRs (Signal-to-Noise Ratios) larger than 10 as PSF star candidates. Our star stamps are arranged into $80\times80$ pixels ($\sim 2.48^{\prime\prime}\times2.48^{\prime\prime}$ for short wavelength channel and $\sim5.04^{\prime\prime}\times 5.04^{\prime\prime}$ for long wavelength channel), and with the result that $\sim40\%$ of the star images have SNR $<40$. Despite the low SNR being inadequate for typical PSF modeling, the central region of these images still offers valuable information for HybPSF due to its emphasis on discrepancies between data and WebbPSF. Finally, we further preserve candidates with ellipticity within 3$\sigma$ to exclude the potential contaminated images. Figure \ref{star_stamps} shows examples of star stamp candidates and the final selected star images of $jw02736001001\_02101\_00002\_nrcalong\_cal.fits$, which is the second exposure of Module A in long wavelength channel ``F277W". Processed images show that the contamination at the margin of the stamps is masked effectively when compared with the candidate images, and significant diffraction structures are visible in the star images.

\begin{figure}[ht!]
\centering
\includegraphics[width=.48\textwidth,clip]{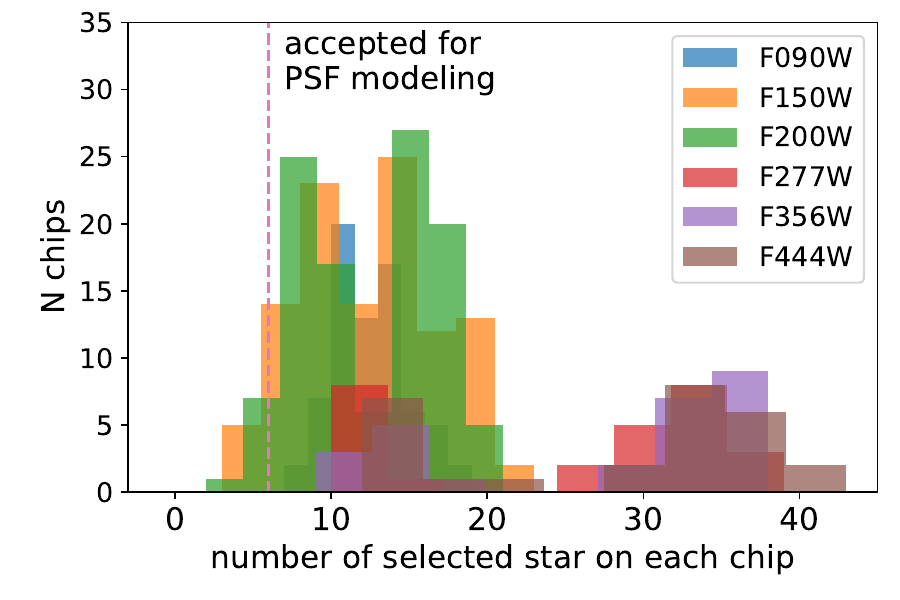}
\caption{Distributions of our select PSF star numbers on single exposed chip for each channel, respectively. The red dashed lines represent the threshold number of stars for PSF modeling in this work. Chips with a selected star number larger than 6 are accepted for PSF modeling and are marked as "accepted for PSF modeling" on the right-hand side of the image.}
\label{snumber_chip}
\end{figure}

Figure \ref{snumber_chip} shows  the number of selected stars on each chip from all of the NIRCam Imaging data of SMACS0723 used in this work. The histogram of star numbers shows two peaks, corresponding to the short and long wavelength filters, respectively. On average, approximately 12 stars per chip are detected from the short wavelength band data, which is significantly lower than the typical number (~100) required for precise PSF modeling \citep{2009A&A...500..647P,2008A&A...484...67P}, and we will discuss later on the consequences of this low number on the analysis.

\begin{figure*}
\centering
\includegraphics[width=1.\textwidth,clip]{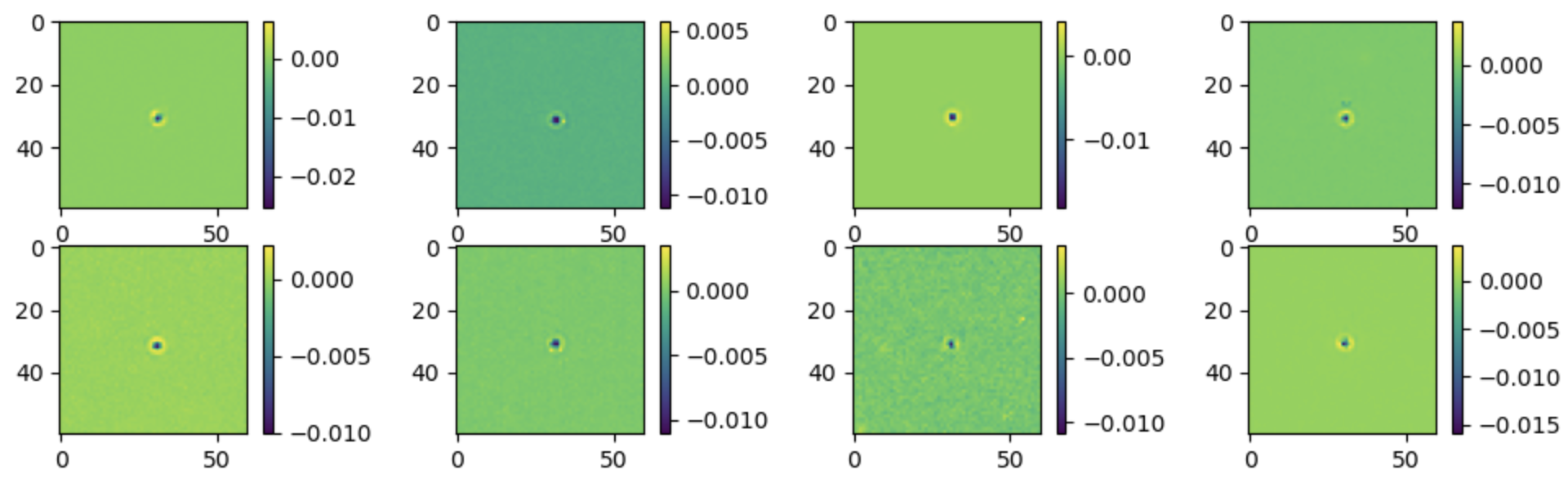}
\caption{Residual image samples of the star images after subtracting the WebbPSF PSF images.}
\label{fig:residual}
\end{figure*}

\begin{figure}[ht!]
\centering
\includegraphics[width=.48\textwidth,clip]{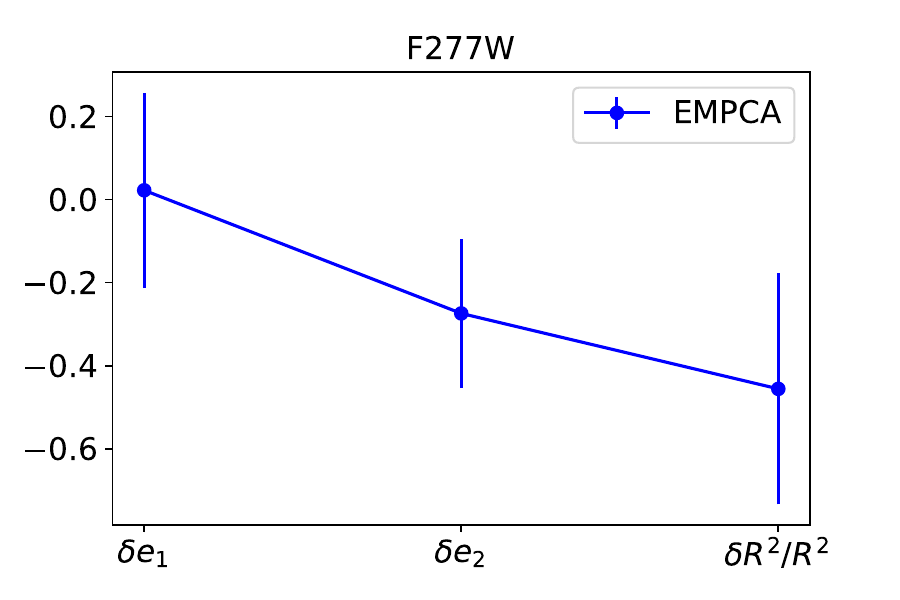}
\caption{Statistical results of the shape parameter residuals (${\delta R^2}/{R^2}$,$\delta e_1$ and $\delta e_2$) for the channel ``F277W'' from EMPCA method.}
\label{emcpa}
\end{figure}

\section{discrepancy between WebbPSF PSF model and observed data }
\label{sec:discrepancy}
WebbPSF provides an easy-to-use way to generate accurate PSF for JWST NIRCam images. We first check the consistency of the generated PSFs with the observed data by generating PSFs with WebbPSF using the corresponding positions, detectors, and filters. In addition, the telescope Optical Path Difference (OPD) files are used, and the OPD information is loaded through the ``$load\_wss\_opd\_by\_date$" method\footnote[4]{\url{https://webbpsf.readthedocs.io/en/latest/jwst_measured_opds.html}} in WebbPSF, where the argument of ``DATE-BEG'' can be read from the header of the corresponding fits image.

We compared the profiles of WebbPSF PSFs and star images. The center of the WebbPSF model is generated at the center of a pixel, while the center of the star images is randomly sampled within a pixel. To make a fair comparison, we first generated the PSFs of corresponding star images with an "oversample" factor of 2. Then, we aligned the centers of the oversampled PSFs with the star images using cubic-spline interpolation. The centers of the star images were estimated using the fast shape estimation algorithm \citep{2012arXiv1203.0571L}. Next, we downsampled the oversampled PSF images to the same resolution as the data. Finally, we masked both the star images and the corresponding PSFs to have the same masked area, as mentioned in section \ref{sec:image}, and normalized their fluxes.

Figure \ref{fig:residual} shows the residual images obtained by subtracting the corresponding PSFs from the star images, which were selected from $jw02736001001\_02101\_00002\_nrcalong\_cal.fits$. The residuals exhibit similar patterns, with almost all of them having negative values in the central region. This indicates that the profile of the simulated PSF by WebbPSF is sharper than that of the star images. This discrepancy has also been reported by \cite{2022ApJ...939L..28D} and \cite{2022arXiv220813582O}. The difference in value of some residuals could be larger than 0.02. Such a remarkable difference between the PSF model and the actual PSFs could introduce systematic errors into the final measurements of astrometry, photometry, and cluster mass reconstruction.

The general way to construct the PSF model for a galaxy is to first reconstruct the PSF using the observed star images, which represent the exact PSFs at the star positions, and then interpolate the PSF to the target positions in the same image, due to the PSF variations in the field. The number of stars required depends on the different interpolation methods \citep{2012MNRAS.419.2356B,2013A&A...549A...1G,2020A&A...636A..78S}. Usually, the PSF reconstruction is applied on a single Charge-Coupled-Device (CCD) chip due to the potential discontinuities of the PSF field across the chips on single exposures \citep{dahle2002weak,2009AAS...21346026J,2011PASP..123..596J,2013ApJ...765...74J,2021ApJ...918...72F,2022MNRAS.517..484N}. This limits the number of stars available for PSF modeling. Moreover, the star images have very complex structures, as shown in Figure \ref{star_stamps}, which increases the complexity \citep{2008A&A...484...67P} of the PSF model. Constructing a PSF model with high complexity requires numerous stars \citep{2008A&A...484...67P,2009A&A...500..647P}. However, as shown in Figure \ref{snumber_chip}, the number of selected stars is fairly low, especially for the short wavelength band, which may not be enough to construct a high-precision PSF model.

Principal component analysis (PCA), a non-parametric data compression method, is widely used in PSF modeling for HST \citep{2007PASP..119.1403J,2009ApJ...697.1793N,2010A&A...516A..63S,2017ApJ...851...46F,2023ApJ...942...23F,2023arXiv230107725F}. However, the star images are too noisy to construct accurate PSF modeling using the PCA method. Expectation-Maximization-PCA (EMPCA) \citep{2012PASP..124.1015B}, an improved PCA method, provides an approach to compose Principal-components (PC) with noisy data-sets or with missing values. \cite{2016RAA....16..139L} used EMPCA to reconstruct the PSF with high accuracy, and \cite{2017MNRAS.471..750M} modeled optical images of galaxies with high precision using the EMPCA method. This suggests that EMPCA might also be suitable for our star images with masked pixels and low SNRs. In the next step, we use the long wavelength data of channel "F277W" to test the PSF reconstruction efficiency of the EMPCA method.
To evaluate the reconstruction effectiveness of EMPCA, we assess the size $R$ and ellipticity $e$ \citep{2008A&A...484...67P,2009A&A...500..647P} of the PSFs. These parameters are used to estimate the potential errors in the PSF model constructed by EMPCA. $e$ and $R$ are usually defined based on the central second brightness moments $Q_{ij}$ of the PSF profile \citep{1998ApJ...504..636H,2001PhR...340..291B}:
\begin{eqnarray}
e_1=\frac{Q_{11}-Q_{22}}{Q_{11}+Q_{22}},\ e_2=\frac{2Q_{12}}{Q_{11}+Q_{22}},\rm{and}\ R^2=Q_{11}+Q_{22}.\nonumber \\
\label{e,r}
\end{eqnarray} 

To suppress the noise at the outskirts of the stamps for the $e$ and $R$ measurements, we use a circular Gaussian function with a full-width at half-maximum (FWHM) of approximately 9 pixels.

Figure \ref{emcpa} shows the shape residuals between the data and reconstructed PSFs from EMPCA. The discrepancy between the model and the data could be 0.4 or more, which is mainly caused by the low SNR and limited number of star images used for PSF modeling. The results are far from the general reconstruction accuracy ($\sim$0.01) required for cluster mass by weak lensing \citep{2005ApJ...618...46J,2009A&A...504....1H}. Therefore, the customary PSF modeling methods (such as EMPCA) might be inadequate for this situation. It is worth mentioning that \cite{2023arXiv230402054F} obtained a very excellent PSF modeling accuracy based on the coadded image of channel F200W. The primary difference between our results and theirs is that they utilized star images with a relatively high SNR of 80$\sim$1000 to construct the inner region (with a stamp size of approximately $0.62^{\prime\prime}\times0.62^{\prime\prime}$ in the channel of F200W) of the PSFs.

Although there are some differences between the PSF model from WebbPSF and the observed data, the models still display excellent agreement with the data. Furthermore, we can see the prominent diffraction spike structures \citep{2022ApJ...936L..14P}, which can also be found in the WebbPSF PSF model \citep{2012SPIE.8442E..3DP}, in the observed data. However, this structure is hard to reconstruct from the PSF star images because star images are fairly faint, and the spikes are totally drowned in noise. Thus, WebbPSF provides us invaluable information about the PSF outer region profile for JWST NIRCam imaging. And the similarity of the residuals gives us a potential possibility to improve the WebbPSF PSF model. We will introduce our supplementary method for improving the WebbPSF model in the next section.

\section{Hybird PSF models by combining WebbPSF and residuals}
\label{sec:hybrid}

In this paper, we attempt to improve the WebbPSF PSF model by addressing the residuals. As discussed earlier, the key question now is how to parameterize the residual images. Given the similar structures of the residual images, we can use the PCA method to extract the most dominant structures of residuals. In the PCA scenario, the approach can be expressed as follows:
\begin{eqnarray}
I_i&=&m_iWebbPSF_i+Res_i, \nonumber \\
&=&m_iWebbPSF_i+\Sigma_{l=1}^{l_{\rm{max}}} P_lC_{il}+noise. 
\label{eq:1}
\end{eqnarray}
where $I_i$ represents the $i$th star image in the field. $WebbPSF_i$ is the corresponding WebbPSF PSF model, and $m_i$ is provided for WebbPSF to capture the potential variations in the field. 
$Res_i$ is the corresponding residual image, $P_l$ represents the $l$th principal component, and $C_{il}$ is the coefficient of $P_l$.
Finally, the hybrid-PSF(HybPSF) is defined as:
\begin{eqnarray}
\rm{HybPSF}_i&=&m_iWebbPSF_i+\Sigma_{l=1}^{l_{\rm{max}}} P_lC_{il}. 
\label{eq:2}
\end{eqnarray}
And the number of principal components $l$ used in this work is $l_{max}=10$, which provides the best trade-off between time consumption and reconstruction performance for the entire residual dataset.

In the classical PCA scenario, the star images should be center-aligned, which is usually achieved through interpolation \citep{2007PASP..119.1403J,2011PASP..123..596J}. However, interpolation can cause potential problems such as aliasing effects and smoothing. To avoid these issues, we use a modified iSPCA method \citep{2021MNRAS.503.4436N} to construct the principal components of the residual images. iSPCA uses a set of oversampled basis functions to align the center of observed images without modifying the data. The principal components are then composed of these basis functions:  $P_l=\Sigma_{m=1}^{m_{\rm{max}}}D_mB_m$, where $B_m$ is the basis function and $D_m$ is the coefficients of the basis functions. In \cite{2021MNRAS.503.4436N}, the Moffatlets basis function showed flexibility in capturing complex structures, so we also use the Moffatlets basis function in this work.
Because we want to construct the principal components for the residual images rather than the star images, the $\chi^2$ equation of iSPCA can be rewritten as:
\begin{eqnarray}
\chi^2&=&\sum_{i=1}^{N_{\rm star}}\sum_{k=1}^{N_{\rm pixel}}\left({I_{ik}-\rm{HybPSF}_{ik}}\right)^2{{W_{ik}}} ,\nonumber \\
&=&\sum_{i=1}^{N_{\rm star}}\sum_{k=1}^{N_{\rm pixel}}\left({I_{ik}-m_iWebbPSF_{ik}-\Sigma_{l=1}^{l_{\rm{max}}} P_lC_{il}}\right)^2{{W_{ik}}}. \nonumber \\
\label{eq:chi}
\end{eqnarray}

In equation \ref{eq:chi}, $W_{ik}$ represents the weight of the $ik$th pixel, where the subscript $l$ and $k$ represent the order of basis functions and pixel index, respectively. In our approach, $W_{ik}$ is the inverse of the estimated noise variance, defined as $\frac{1}{\sigma_{ik}^2}$ \citep{2021MNRAS.503.4436N}. However, in our case, the star images have masked pixels that need to be considered. Therefore, we set ${W_{ik}}=\frac{\mathcal{M}}{\sigma_{ik}^2}$, where $\mathcal{M}$ is the mask image mentioned in section \ref{sec:image}.

The basis function in equation (\ref{eq:chi}) can be sampled at any resolution theoretically. However, in this work, we consider the time consumption and the oversampling factor of the WebbPSF PSF model. Therefore, we generate the basis functions with an oversampling factor of $``oversample''\times2$. This means that the constructed principal components are also oversampled. We then downsample the principal components to the same resolution as the WebbPSF PSF model to construct the HybPSF.

\begin{figure*}[t]
\centering
\includegraphics[width=1\textwidth,clip]{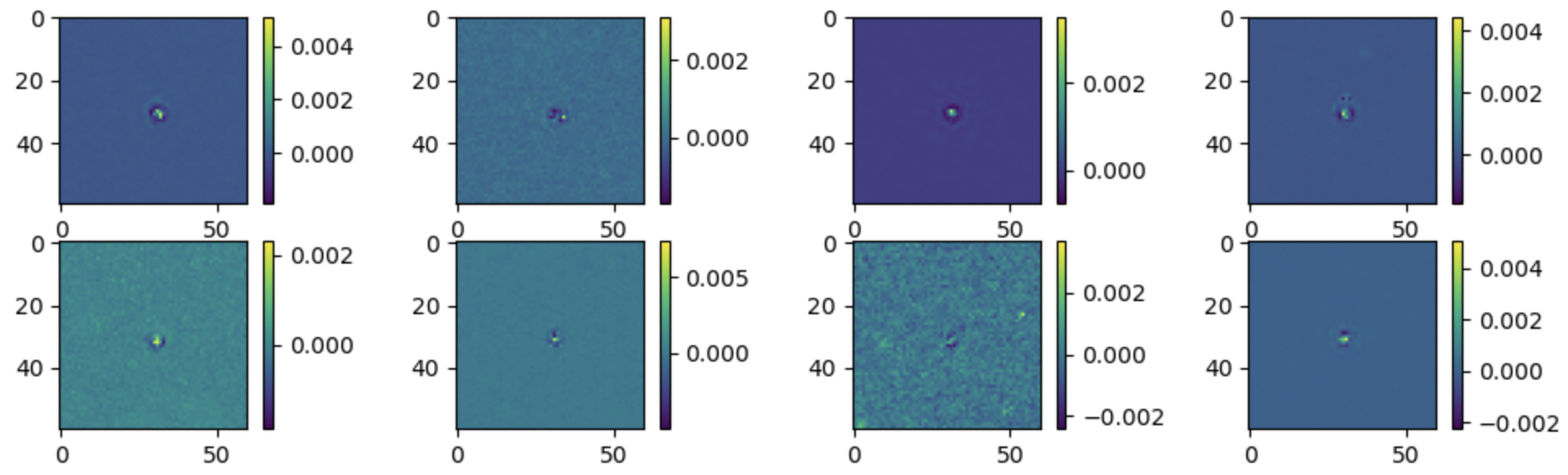}
\caption{Residual image samples of the star images after subtracting the HybPSF model images. }
\label{fig:hresidual}
\end{figure*}

\begin{figure*}[t]
\centering
\includegraphics[width=.9\textwidth,clip]{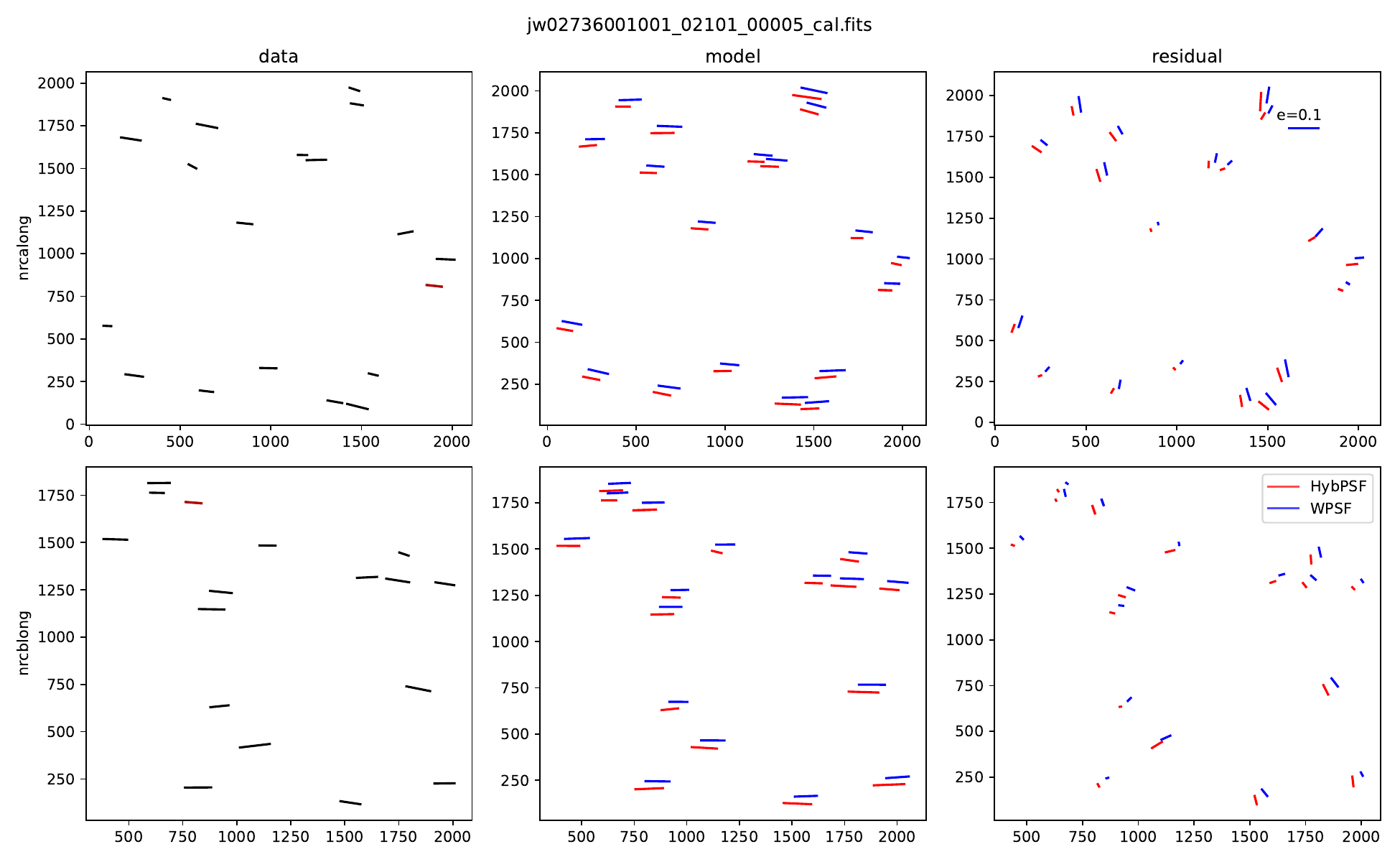}
\caption{Ellipticity comparison between the data and the modeled PSFs in channel F277W. The whiskers mimic the elliptictiy of the PSFs.The left, middle and right panel correspond to the ellipticity of star image, model and residual subtracting the model from data respectively.}
\label{fig:ecom}
\end{figure*}

\begin{figure*}[t]
\centering
\includegraphics[width=.7\textwidth,clip]{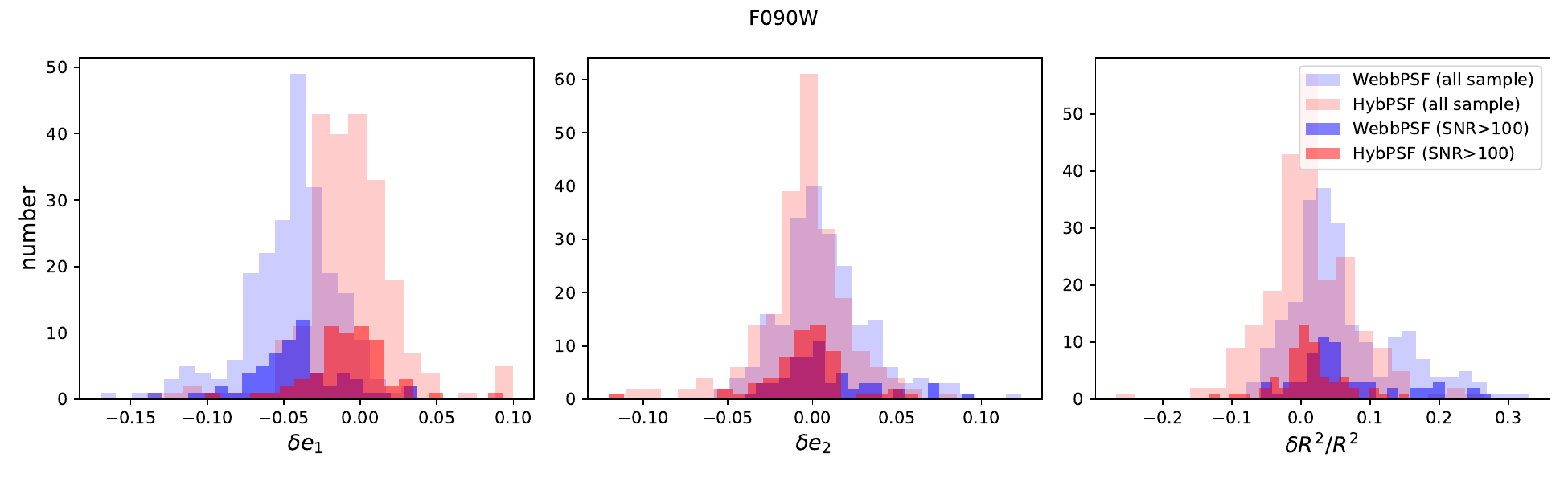}
\includegraphics[width=.7\textwidth,clip]{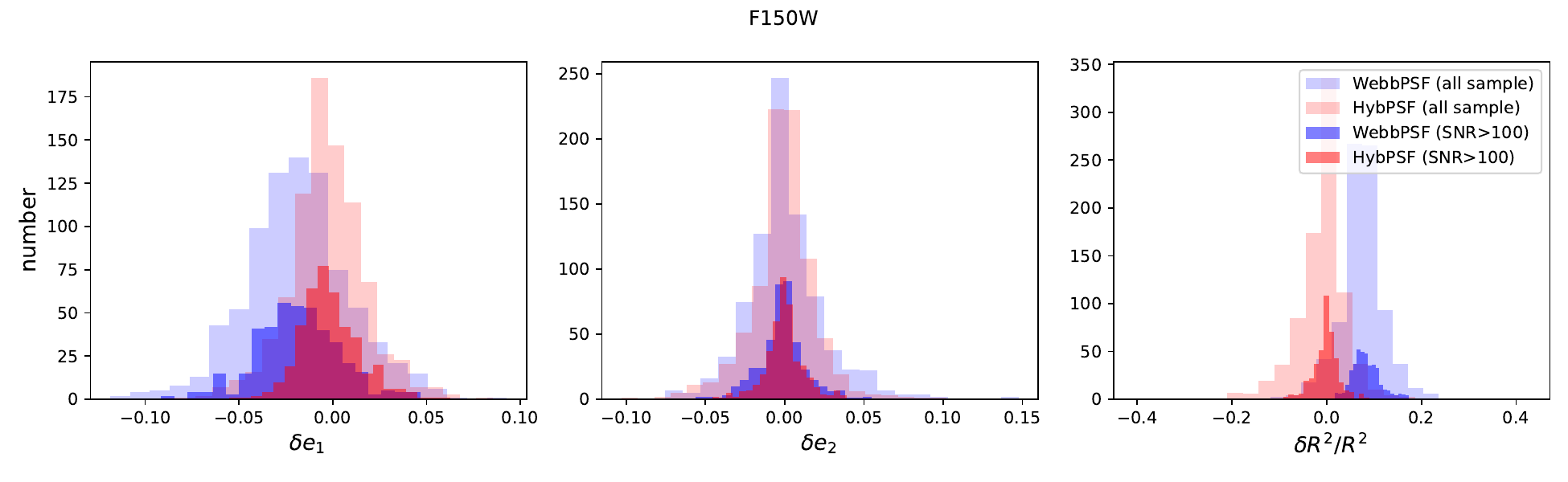}\\
\includegraphics[width=.7\textwidth,clip]{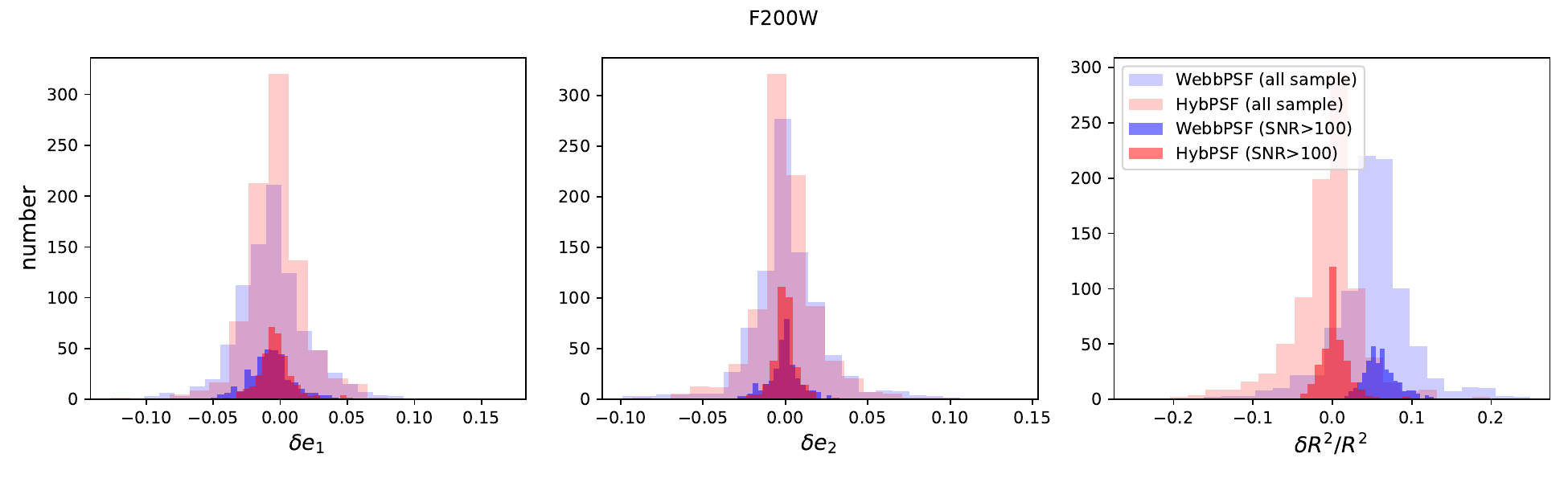}
\includegraphics[width=.7\textwidth,clip]{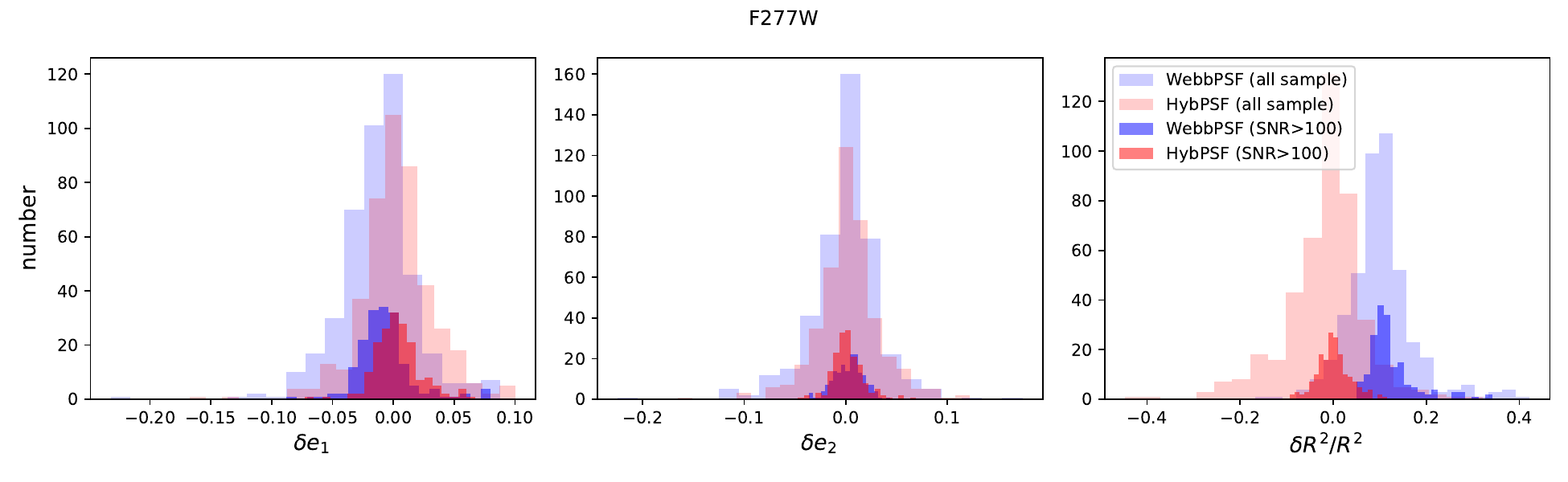}\\
\includegraphics[width=.7\textwidth,clip]{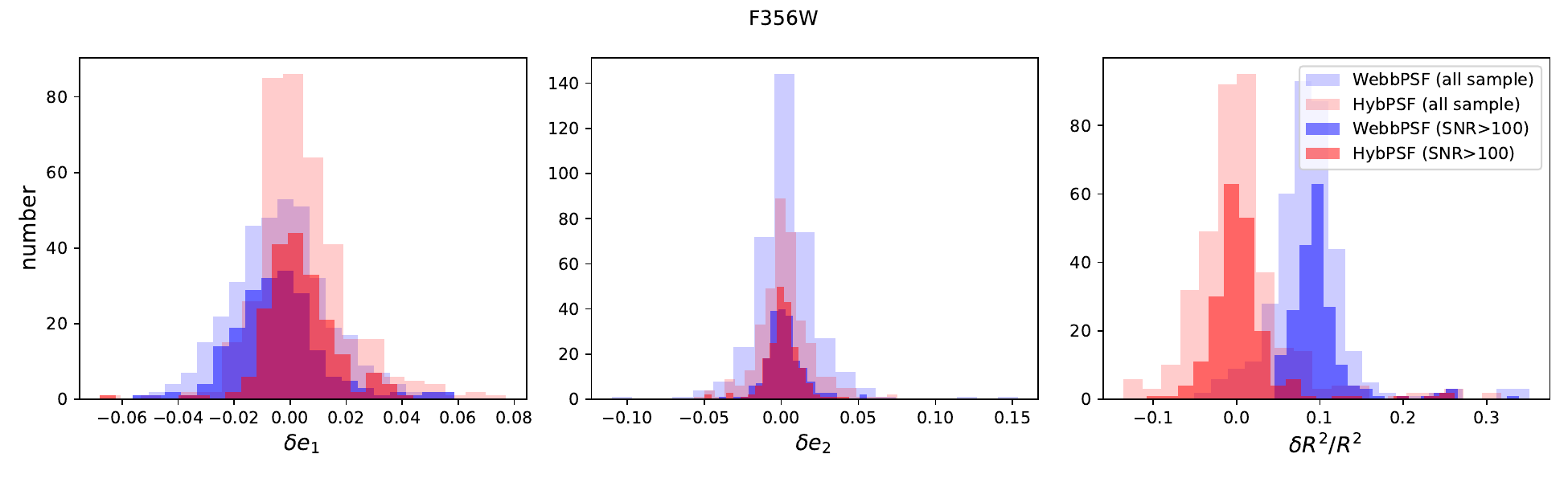}
\includegraphics[width=.7\textwidth,clip]{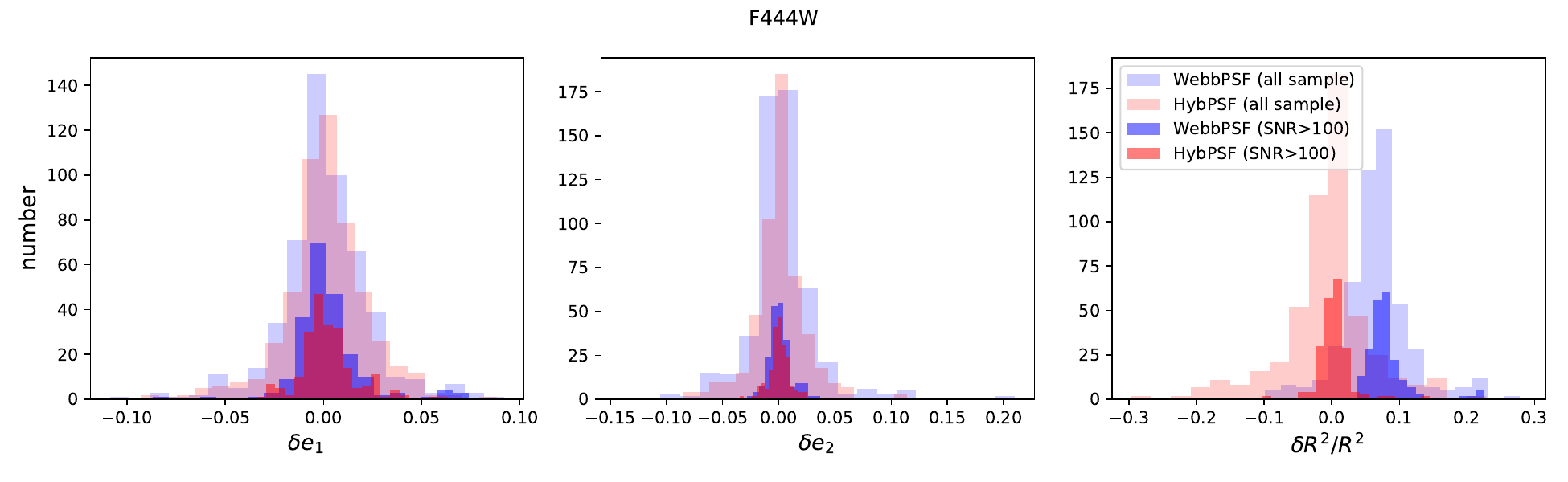}
\caption{Distribution of the shape parameter residuals (${\delta R^2}/{R^2}$,$\delta e_1$ and $\delta e_2$) for the JWST NIRCam imaging data used in this work. The blue and red histogram represent the results obtained from WebbPSF and HybPSF, respectively. Lighter color histograms represent the reuslts from all samples.}
\label{fig:distr}
\end{figure*}

It is worth noting that although principal components are used to construct the residuals, the basis functions must be aligned with the center of star images. Additionally, the residuals exhibit the most significant discrepancy in the central region of the images, so we can only calculate the $\chi^2$ in the inner region of the stamps. For this work, we used only the central $44\times 44$ pixels of each image to calculate $\chi^2$, rather than the entire stamp area, which also helped to reduce time consumption. After calculating the coefficients of $m_i$, $C_{il}$ and $D_{im}$ from equation \ref{eq:chi}, we reconstructed the PSF of each star image by combining the $WebbPSF_i$ and $Res_i$ models from iSPCA, respectively. In order to capture the spatial variations of the PSFs, we used bivariate polynomials to fit the coefficients of the principal components and WebbPSF in the field for each CCD chip \citep{2007PASP..119.1403J}. We found that polynomials below second order were unable to fit the residuals field well, so we only adopted chips with PSF star numbers larger than 6 for PCA. Figure \ref{snumber_chip} shows the distribution of star numbers on the accepted chips on the right side of the pink dashed line, while chips on the left were not used in this work.

Now we summarize the model process of HybPSF as follows :

\uppercase\expandafter{1}. $Extracting\ star\ images$: In this step, we extract the PSF star images and construct the masked areas and weight images and then normalise each images.

\uppercase\expandafter{2}. $Constructing\ residual\ images$: Generating the corresponding WebbPSF, and subtracting the centrally aligned and flux normalised model from the star images to get the residual images.

\uppercase\expandafter{3}. $Calculating\ \chi^2\ equation$: Aligning the WebbPSF to the center of each data respectively, and use the data, WebbPSF, and masked images to calculate the $\chi^2$ equation. 

\uppercase\expandafter{4}. $Constructing\ the\  {\rm HybPSF}\ model$: Applying the polynomial fitting to the coefficients $C_{il}$ and $m_i$ of each chip respectively, and combining the WebbPSF and $Resi$ model at target positions from WebbPSF and fitted polynomial respectively. The maximum order of polynomials depends on the number of PSF stars used in the PSF reconstruction, for example: when the number of PSF stars is greater than 21, the maximum order is 5.

Figure \ref{fig:hresidual} displays the residual image examples of the star images after subtracting the HybPSF, which correspond to $jw02736001001\_02101\_00002\_nrcalong\_cal.fits$. We observe that the maximum difference between HybPSF and the data is less than 0.006, which is nearly one order of magnitude smaller than that shown in Figure \ref{fig:residual}. The pattern of the residual images shows less similarity, indicating that HybPSF has successfully extracted the most prominent structure in the residuals. However, the residuals still exhibit some structures, which may result from the small number of stars used for PCA decomposition. Overall, almost all of the images show considerable improvement. Figure \ref{fig:ecom} provides an example of ellipticity comparison between the data, WebbPSF, and HybPSF. The primary features of the ellipticity distributions in the field are captured by HybPSF and WebbPSF, and the distribution of whiskers in the residual panel appears random.

\begin{figure*}[t]
\centering
\includegraphics[width=.3\textwidth,clip]{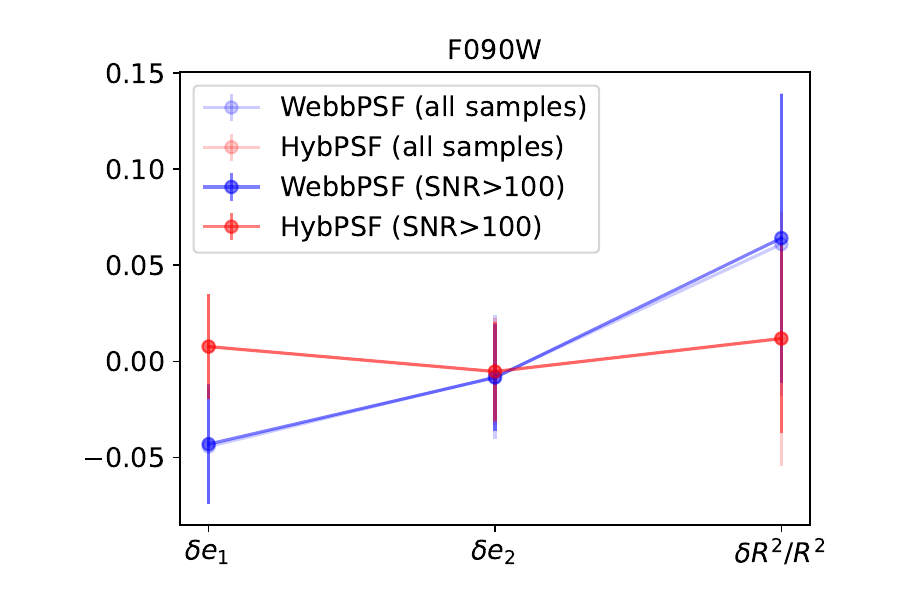}
\includegraphics[width=.3\textwidth,clip]{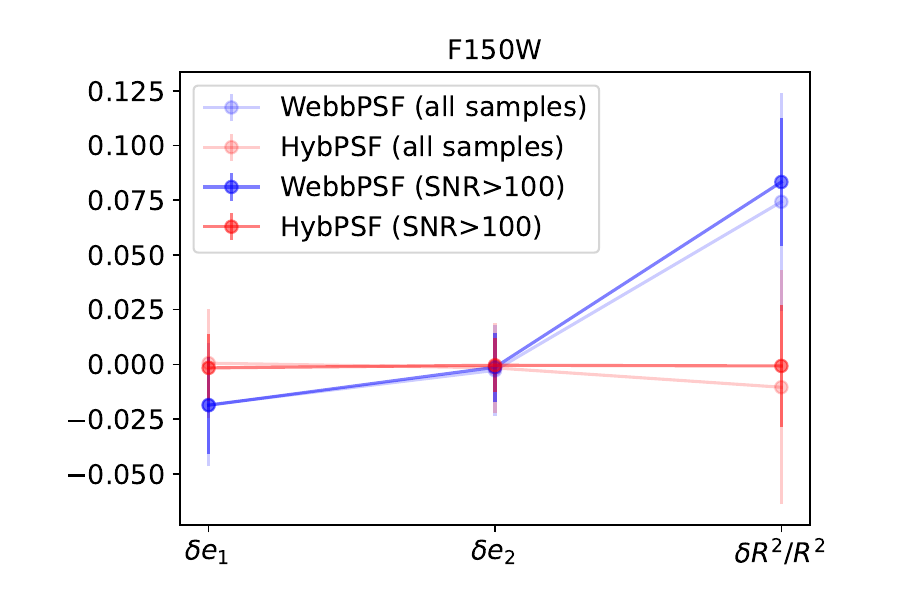}
\includegraphics[width=.3\textwidth,clip]{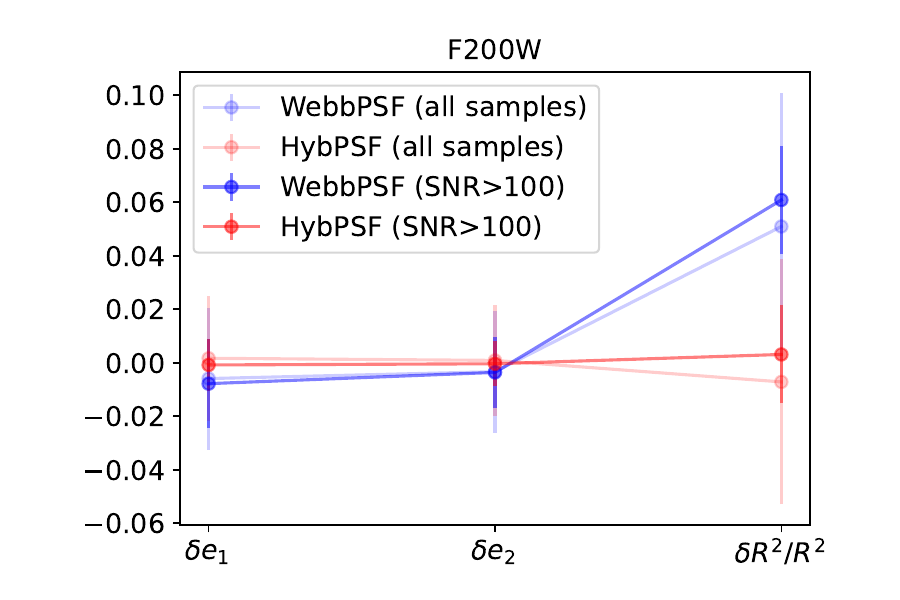}\\
\includegraphics[width=.3\textwidth,clip]{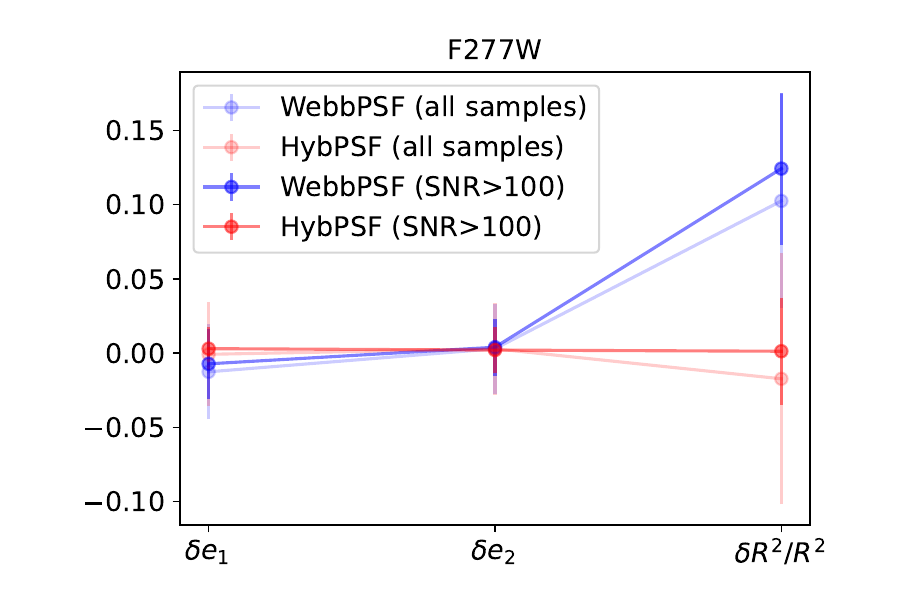}
\includegraphics[width=.3\textwidth,clip]{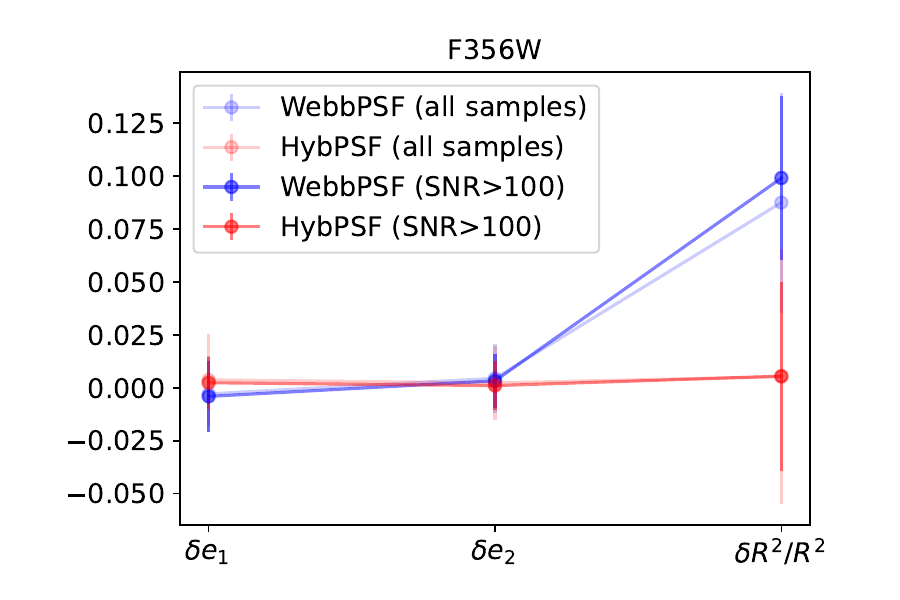}
\includegraphics[width=.3\textwidth,clip]{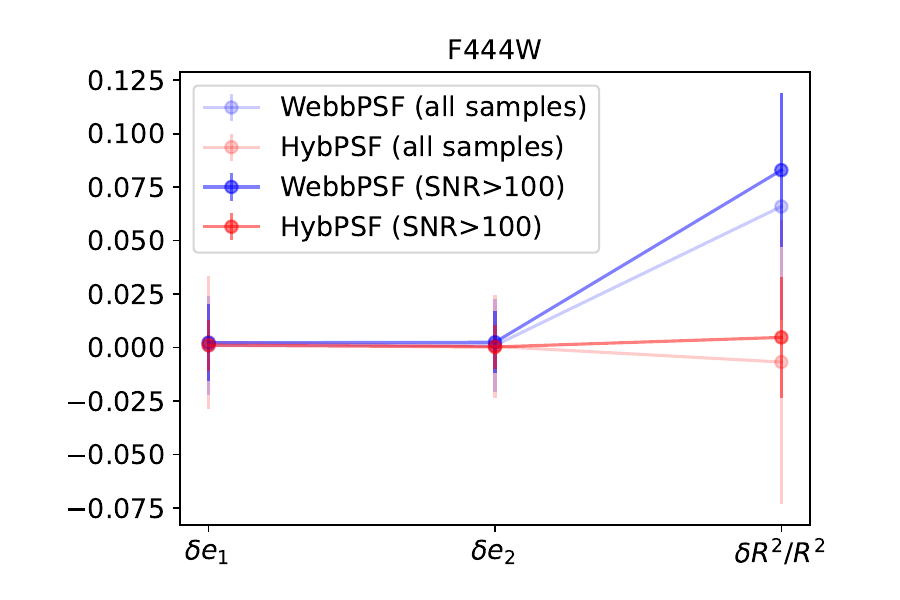}
\caption{Statistical results of the shape parameter residuals (${\delta R^2}/{R^2}$,$\delta e_1$ and $\delta e_2$) for the JWST NIRCam imaging data used in this work. The blue and red points represent the results obtained from WebbPSF and HybPSF, respectively. The error bars represent the 1$\sigma$ width of the distributions.}
\label{fig:shape}
\end{figure*}

Figure \ref{fig:distr} shows distributions of the shape parameter residuals (${\delta R^2}/{R^2}$,$\delta e_1$ and $\delta e_2$) for the JWST NIRCam imaging data used in this work, respectively. And $\delta e$ and $\delta R^2/R^2$ values are obtained by subtracting the $e$ and $R$ of the PSF model from the measurement of the star images. The light-colored distributions in Figure \ref{fig:distr} represent the results from all of the samples, including images with low SNRs, which are used as a contrast. As shown in Figure \ref{fig:distr}, the ${\delta R^2}/{R^2}$ distributions in channels F150W, F200W, F277W, F356W, and F444W, and the $\delta e_1$ distributions in F090W appear separated for the two methods. However, all of the residuals are spread over a wide range for both methods. To avoid the noise effect in shape estimations, we use star images with high SNRs (SNR$>$100) as a contrast to obtain more explicit comparisons. The results, shown in Figure \ref{fig:distr} with a heavier color, demonstrate that the ${\delta R^2}/{R^2}$ and $\delta e_1$ distributions for the two methods mentioned above are clearly separated. The peaks of the HybPSF distribution (shown in red) are closer to 0 compared to those of the WebbPSF distribution (shown in blue). The results from high SNR data indicate more compact distributions.

Furthermore, we present the statistical results of $\delta e$ and $\delta R^2/R^2$ for the two methods in Figure \ref{fig:shape}. As with Figure \ref{fig:distr}, the results are obtained for two types of datasets (all samples and high SNR samples). The scatter plots from high SNR data show much smaller ranges than those from all samples. The results of  $\delta R^2/R^2$ are clearly separated, suggesting that the size precision of the PSF model is improved significantly, and the ellipticity precision is also improved to some extent. Table \ref{tab:shape} shows the mean shape residuals for WebbPSF and HybPSF in each channel from samples with high SNRs, respectively. The results of HybPSF are closer to zero than that of WebbpSF. Specifically, the mean residuals of $\delta R^2/R^2$ in the F150W, F200W, F277W, F356W, and F444W channels decreased by an order of magnitude (from a few percent to parts per thousand). Additionally, the values of the ellipticity component $\delta e_1$ in channel F090W also show a decline. 
The 1$\sigma$ uncertainty of the residuals is constrained to few percent, indicating that the model may be acceptable for cluster mass analysis using weak lensing \citep{2005ApJ...618...46J,2009A&A...504....1H}.

Based on Figures \ref{fig:distr} and \ref{fig:shape}, the mean residual values of HybPSF from all samples and high SNRs samples are statistically consistent. This suggests that low SNRs images could also provide effective constraints for JWST NIRCam imaging PSF reconstruction in our method.

\begin{table*}
	\centering
	\caption{The mean value and 1$\sigma$ width of the residual distributions.
	}
	\label{tab:shape}
	\begin{tabular}{lcccccccccccccr} % four columns, alignment for each
		\hline
            \hline
            Methods & Channels & $<\delta e_1>$ &  1$\sigma$ width & $<\delta e_2>$ & 1$\sigma$ width & $<\delta R^2/R^2>$ & 1$\sigma$ width &  number of contrast \\
		\hline
		& F090W & -0.031 & 0.021 & -0.005 & 0.023 & 0.097 & 0.068 &194  \\
		& F150W & -0.017 & 0.018 & -0.0002 & 0.014 & 0.083 & 0.025 & 334 \\
            WebbPSF & F200W & -0.007 & 0.011 & -0.003 & 0.011 & 0.060 & 0.016 & 317    \\
            & F277W & -0.006 & 0.020 & 0.003 & 0.017 & 0.125 & 0.052 & 172 \\
            & F356W & -0.004 & 0.016 & 0.003 & 0.012 & 0.099 & 0.038 & 200  \\
            & F444W & 0.002 & 0.016 & 0.002 & 0.011 & 0.083 & 0.035 & 212 \\
            \hline
            & F090W & 0.001 & 0.015 & -0.001 & 0.011 & 0.024 & 0.051 & 194\\
            & F150W &  0.0005 & 0.016 & 0.0004 & 0.013 & -0.000 & 0.048  & 334\\
            HybPSF& F200W & -0.001 & 0.007 & -0.0002 & 0.004 & 0.001 & 0.014  & 336 \\
            & F277W & 0.003 & 0.012 & 0.001 & 0.012 & 0.001 & 0.036 & 172\\
            & F356W & 0.002 & 0.012 & 0.001 &  0.011 & 0.005 & 0.044 & 200 \\
            & F444W & 0.001 & 0.010 & 0.0003 & 0.006 & -0.004 & 0.027 & 212 \\
            \hline
            \hline
	\end{tabular}
\end{table*}

The HybPSF model can be regarded as the central profile calibrated WebbPSF, and the ellipticity has more dependence on the peripheral profile of the PSF according to its definition. Hence, the improvement for PSF size is more evident than that for ellipticity. We also observe that the error of $\delta e$ is smaller than that of $\delta R^2/R^2$ in Figure \ref{fig:shape}.
Our method can also reconstruct the PSFs with high precision for the "stage three" data products of JWST, which may be used more widely. To demonstrate this effect, we use the Drizzle coadded images \citep{2002PASP..114..144F,2017RAA....17..100W} from one chip in the long wavelength channels for PSF reconstruction using HybPSF, since they would expect to have more PSF stars with high SNRs. In the coadded images, we choose star images with  SNR$>$40 for PSF modeling and images with SNR$>$100 for contrast. Additionally, we restrict the comparison region of the PSFs to the inner $30\times30$ pixels to avoid noise effects. Generally, the stacked image would not be the average of multiple exposures due to rotation, geometric distortion, etc. \citep{2002PASP..114..144F}. For simplicity, we construct the WebbPSF for coadded images by averaging the WebbPSF corresponding to multiple exposures obtained for each star.

\begin{figure*}[t]
\centering
\includegraphics[width=.3\textwidth,clip]{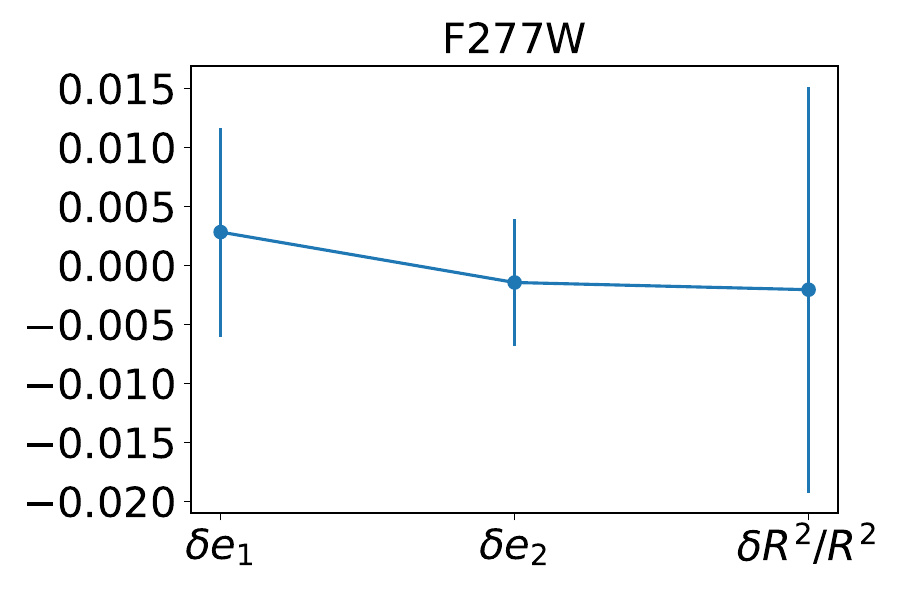}
\includegraphics[width=.3\textwidth,clip]{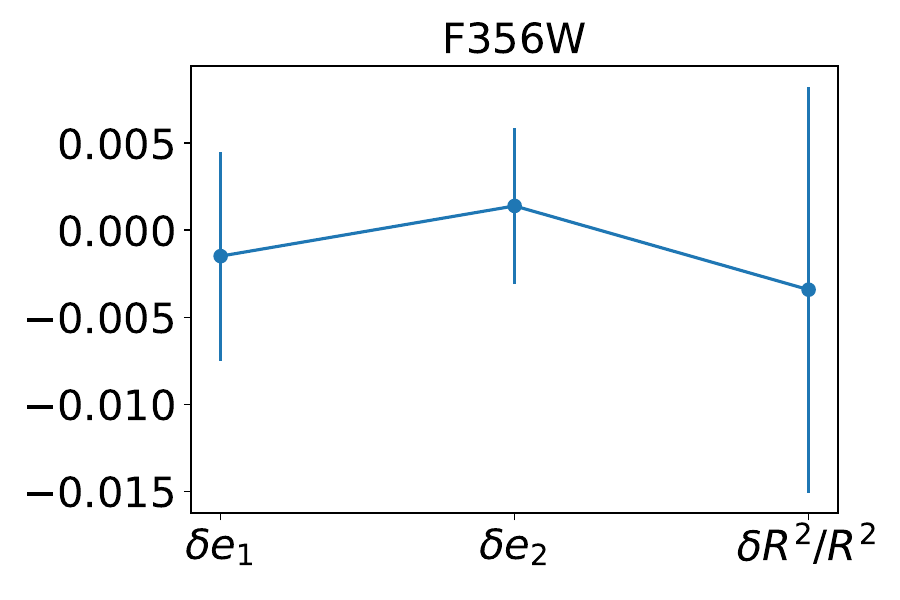}
\includegraphics[width=.3\textwidth,clip]{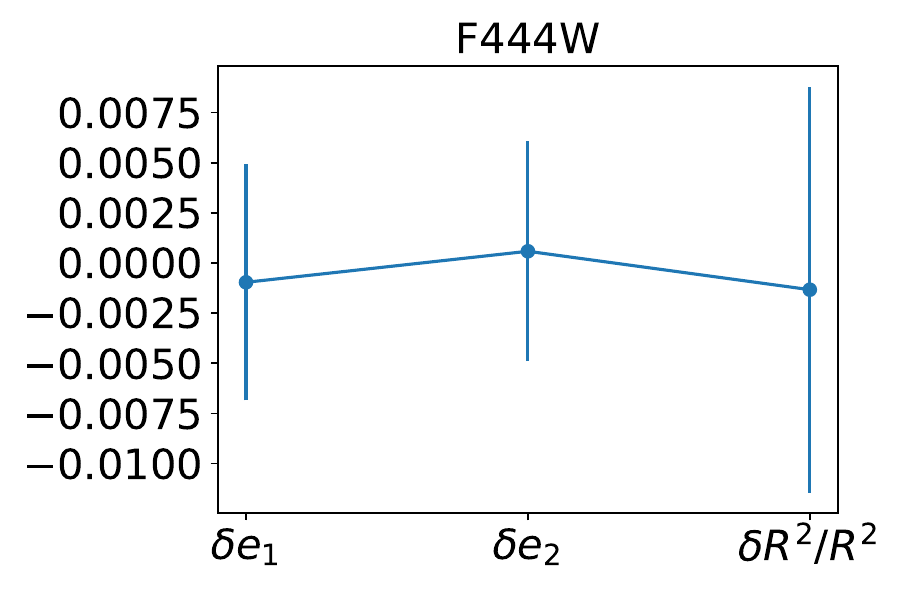}
\caption{Statistical results of the shape parameter residuals estimated based on coadded images (${\delta R^2}/{R^2}$,$\delta e_1$ and $\delta e_2$) of channel F277W, F356W and F444W, respectively.  The error bars represent the 1$\sigma$ width of the distributions.}
\label{fig:coshape}
\end{figure*}

\begin{table*}
	\centering
	\caption{The mean value and 1$\sigma$ width of the residual distributions for coadded data.
	}
	\label{tab:coshape}
	\begin{tabular}{lcccccccr} % four columns, alignment for each
		\hline
            \hline
            {} & Channels & $<\delta e_1>$ & 1$\sigma$ width & $<\delta e_2>$ & 1$\sigma$ width & $<\delta R^2/R^2>$ & 1$\sigma$ width & number of contrast  \\
		\hline
             & F277W & 0.002 & 0.008 & -0.001 & 0.005 & -0.002 & 0.017 & 27  \\
            HybPSF & F356W & -0.001 & 0.006 & 0.001 & 0.004 & -0.003 & 0.011 & 23 \\
             & F444W & -0.0009 & 0.005 & 0.0005 & 0.005 & -0.001 & 0.010 & 23 \\
		\hline
            \hline
	\end{tabular}
\end{table*}

Figure \ref{fig:coshape} displays the shape residual results of the coadded images, and the width of the residual distributions is much smaller than those in Figure \ref{fig:shape}. The 1$\sigma$ uncertainty of these points is markedly reduced to $\sim$0.01 for $\delta R^2/R^2$ and parts per thousand for  $\delta e$, as shown in Table \ref{tab:coshape}.

Based on the residuals and shape parameter comparisons, HybPSF appears to have better performance than WebbPSF. HybPSF constructs a residual model based on PCA and provides supplements to the WebbPSF PSF model. We expect this could help us to make better measurements for cluster mass, astrometry, photometry, etc. Furthermore, HybPSF can provide a PSF model with a considerable area that is difficult to achieve using only the empirical method based on star images. This advantage is promising for the upsampling deconvolution method \citep{2022MNRAS.517..787W}. However, we also observe some structures left in the residuals, as seen in Figure \ref{fig:residual}. Therefore, the PSF is not reconstructed perfectly by our method, which may be caused by the small number of star images in the single exposure. Additionally, we did not consider the spectral energy distribution (SED) dependence of each stellar object in WebbPSF, which may help to calculate more accurate PSF models.

\begin{figure*}[t]
\centering
\includegraphics[width=1\textwidth,clip]{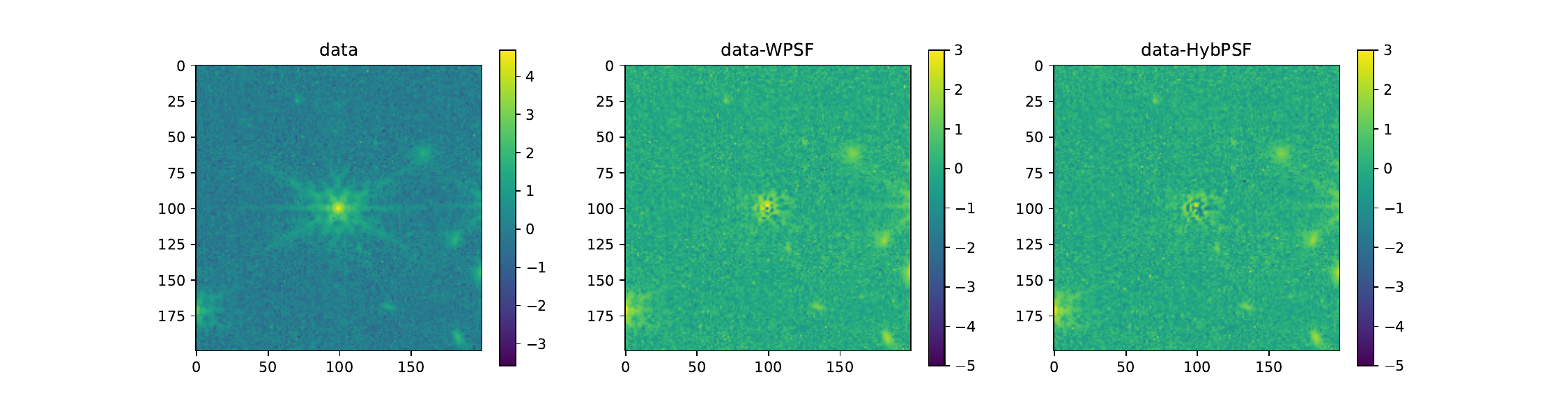}
\caption{Star spike deduction test. $Left$, the observed data. $Middel$, the residual obtained by subtracting WebbPSF from the data. $Right$, the residual obtained by subtracting HybPSF from the data. The images are shown in log scale.}
\label{fig:spike}
\end{figure*}

The shape comparisons above mainly focus on the inner region of PSFs to avoid noise effects. One of the features of HybPSF is that it includes outer profiles, such as the diffraction spikes. We cut a large stamp ($6.3^{\prime\prime}\times 6.3^{\prime\prime}$) with prominent diffraction spikes from the long wavelength channel F356W to test the performance of HybPSF. In this test, we set the coefficient $m$ of WebbPSF to be 1. This is because that the coefficient $m_i$ is estimated from faint star images which show no distinct spikes structures in the very outer regions, and the constraints for considerably bright stars might be inappropriate. Figure \ref{fig:spike} shows the spike correction of WebbPSF and HybPSF, respectively. The star flux is calibrated using the flux of the inner circular region with a radius of 10 pixels from the center of images. The outer region of HybPSF is constructed from WebbPSF, so they have very similar performance, as shown in Figure \ref{fig:spike}. This suggests that HybPSF can also capture the spike structures at very large scales of the JWST NIRCam image PSFs. We then estimate the AB magnitude of the star based on the two PSF models, and the difference in AB magnitude estimated from the two PSF models is $\sim0.01$ magnitude.

\section{SUMMARY} 
\label{sec:summ}

With the powerful observation capability of JWST, we are entering a new era of exploring our universe. However, accurate modeling of its PSF is also required for high-precision astrophysical measurements. In this paper, we introduce a PSF modeling method called HybPSF for the NIRCam images, which combines observed data and simulated PSFs from WebbPSF.

To extract proper star images from the observed data, we began with the "stage two" calibrated file and then filled up the contaminated pixels and performed deblending for each star image in later steps. However, comparing the observed data and the simulated PSF from WebbPSF, we found that the residuals still had significant structures, suggesting that the PSF model could be improved potentially.

We then attempted to extract the remaining signals from the residuals using the iSPCA method, and supplemented the constructed residual model back to WebbPSF to obtain a PSF model (HybPSF) that would be more consistent with the observed data. The value range of the corresponding residuals from HybPSF reveals the improvements of our method. Finally, we checked the model precision on the PSF shape parameters: $e$ and $R$. The $e$ and $R$ of the HybPSF models showed better precision than WebbPSF to some extent, especially on $R$. This suggests that HybPSF is more consistent with the data than WebbPSF. Additionally, our method can provide oversampled PSF images, which could be promising for up-sampling PSF deconvolution (Wang et al. in prep). Additionally, this method is also applicable to i2d.fits images. Given our scientific interest in accurately measuring shear, we aimed to avoid uncertainties in the PSF resulting from image stacking. Therefore, we required accurate PSF information from individual images (cal.fits). The PSF obtained in this work will also be utilized for PSF deconvolution and image stacking (Wang in prep).

However, there are still some discrepancies between our model and the data, which can be seen in the residual images and shape residuals. This may be caused by the limited number of stars used for PSF modeling, and some chips were dropped due to the limited number of stars. This issue may be alleviated by considering all star images in the entire field, as done in \cite{2020arXiv201109835L,2023arXiv230402054F}, and galaxy images may also provide some help in solving this problem \citep{2021MNRAS.508.3785N}. Additionally, SED information of the stellar objects can help WebbPSF construct more accurate PSF models, which we will explore in future work. The brighter-fatter effect is another important systematic in PSF modeling. However, \cite{2023arXiv230402054F} mentioned that this effect would not significantly impact cluster mass reconstruction. We will check for these influences on PSF modeling in the future. Overall, we expect that this improved PSF model will help us obtain better astrophysical measurements, such as cluster mass, photometry, and more.

%% Also note that the akcnowlodgment environment does not support long amounts of text. If you have a lot of people and institutions to acknowledge, do not use this command. Instead, create a new \section{Acknowledgments}.
%\begin{acknowledgments}
\section{Acknowledgments}
This work is supported by National Key R$\&$D Program of China No.
2022YFF0503403 and the Ministry of Science and Technology
of China (grant Nos. 2020SKA0110100). HYS acknowledges the support from NSFC of China under grant 11973070, Key Research Program of Frontier Sciences, CAS, Grant No. ZDBS-LY-7013 and Program of Shanghai Academic/Technology Research Leader. We acknowledge the support from the science research grants from the China Manned Space Project with NO. CMS-CSST-2021-A01, CMS-CSST-2021-A04.

The Early Release Observations and associated materials were developed, executed, and compiled by the ERO production team:  Hannah Braun, Claire Blome, Matthew Brown, Margaret Carruthers, Dan Coe, Joseph DePasquale, Nestor Espinoza, Macarena Garcia Marin, Karl Gordon, Alaina Henry, Leah Hustak, Andi James, Ann Jenkins, Anton Koekemoer, Stephanie LaMassa, David Law, Alexandra Lockwood, Amaya Moro-Martin, Susan Mullally, Alyssa Pagan, Dani Player, Klaus Pontoppidan, Charles Proffitt, Christine Pulliam, Leah Ramsay, Swara Ravindranath, Neill Reid, Massimo Robberto, Elena Sabbi, Leonardo Ubeda. The EROs were also made possible by the foundational efforts and support from the JWST instruments, STScI planning and scheduling, and Data Management teams.

%\end{acknowledgments}



\software{astropy \citep{2013A&A...558A..33A,2018AJ....156..123A},  
          Source Extractor \citep{1996A&AS..117..393B,2010ascl.soft10064B},
          jwst \citep{2017ASPC..512..355B,2019ASPC..523..543B,2022zndo...7038885B},
          WebbPSF \citep{2012SPIE.8442E..3DP,2014SPIE.9143E..3XP,2015ascl.soft04007P}
          }




\bibliography{HPSF}{}
%\bibliographystyle{aasjournal}



\end{document}